\documentstyle[psfig,floats,aps,twocolumn]{revtex}

\begin{document}

\draft

\twocolumn[\hsize\textwidth\columnwidth\hsize\csname %       |
@twocolumnfalse\endcsname

\title{Three-dimensional dendrite tip morphology
at low undercooling}

\author{Alain Karma, Youngyih H. Lee, and Mathis Plapp}

\address{Physics Department and Center for Interdisciplinary Research
on Complex Systems,\\
  Northeastern University, Boston MA 02115}
\date{\today}

\maketitle

\begin{abstract}
We investigate the three-dimensional morphology of the dendrite tip
using the phase-field method. We find that, for low undercoolings,
this morphology is ostensibly independent of anisotropy strength 
except for a localized shape distortion near the tip that
only affects the value of the tip radius $\rho$ (which is crudely
approximated by $\rho\approx
(1-\alpha)\rho_{Iv}$ where $\rho_{Iv}$ is the Ivantsov tip radius
of an isothermal paraboloid with the same tip 
velocity and $\alpha$ is the stiffness anisotropy).
The universal tip shape, which excludes this distortion, 
is well fitted by the form $z=-r^2/2+A_4 r^4\cos 4\phi$ 
where $|z|$ is the distance from the tip and all lengths are scaled 
by $\rho_{Iv}$.
This fit yields $A_4$ in the range $0.004-0.005$ in
good quantitative agreement with the existing tip morphology
measurements in succinonitrile {\rm [}LaCombe {\it et al.}, Phys. Rev. 
E {\bf 52}, 2778 (1995){\rm ]}, which are reanalyzed here and found to be
consistent with a single $\cos 4\phi$ mode non-axisymmetric deviation
from a paraboloid. 
Moreover, the fin shape away from the tip is well fitted by the 
power law $z=-a |x|^{5/3}$ with $a\approx 0.68$. 
Finally, the characterization
of the operating state of the dendrite tip is revisited in the light of
these results.
\end{abstract}
]

\section{Introduction}

The shape of crystal dendrites was first suggested by 
Papapetrou \cite{Pap} to be parabolic, and slightly more
than a decade later Ivantsov demonstrated \cite{Iva} 
that a parabola (paraboloid)
is an exact solution of the steady-state
growth equations in two (three) dimensions 
when capillary effects are entirely neglected
and the interface is isothermal. 
He derived the well-known relationship
\begin{equation}
\Delta=P_{Iv}\exp(P_{Iv})\,\int_{P_{Iv}}^{\infty}
\,ds\,\frac{\exp(-s)}{s}  \label{Ivan}
\end{equation}
between the Peclet number $P_{Iv}=\rho_{Iv} V/(2D)$ and
the dimensionless undercooling $\Delta=(T_M-T_\infty)/(L/c_p)$
for a paraboloid of tip velocity $V$ and tip radius $\rho_{Iv}$ 
where $D$ is the thermal diffusivity,
$T_M$ is the melting temperature, 
$T_{\infty}$ is the initial temperature of the undercooled 
liquid, $L$ is the latent heat of melting,
and $c_p$ is the specific heat at constant pressure.

In more recent history, the development of solvability theory
\cite{Lan:Hou,Kesetal:Rev}
has led to the additional and crucial understanding that the anisotropic 
surface energy acts as a singular perturbation
that uniquely selects the tip velocity, and also alters the
entire dendrite shape \cite{BenBre,Bren2,Bre}.
According to this theory, the scaling parameter
$\sigma^*=2Dd_0/\rho_{Iv}^2V$
(where $d_0$, defined below, is the capillary length)
approaches a constant 
at low undercooling that only depends on the
anisotropy strength (denoted here by
$\epsilon_4$ for a crystal with an underlying cubic
symmetry), and scales as
$\sigma^* \sim \epsilon_4^{7/4}$ in the limit
of vanishingly small anisotropy. In this
same limit, Ben Amar and Brener \cite{BenBre} have predicted that
capillary effects lead to a universal four-fold 
deviation from a paraboloid of the form
\begin{equation}
z=-\frac{r^2}{2}\,+\,A_4 r^4 \cos 4\phi, \label{anishape}\\
\end{equation}
where $A_4=1/88$ is independent of anisotropy strength and
$(r,\phi)$ are the polar coordinates in the plane normal to the
growth axis $z$ with all lengths scaled by $\rho_{Iv}$.
The improved prediction $A_4=1/96$ has been
obtained in a subsequent analysis
\cite{Bren2}. 

To characterize the shape 
further behind the tip, Brener remarked that the
cross-sectional shape of needle crystal dendrites perpendicular
to the growth axis can be assumed to
evolve with increasing distance $|z|$
from the tip as a two-dimensional growth shape in time ($t=|z|/V$).
The area of this shape increases
linearly in time, and thus $|z|$, as
\begin{equation}
S(z)=2\pi \rho_{Iv}|z|\label{area}
\end{equation}
where $\rho_{Iv}$ is defined by Eq. \ref{Ivan} with $V$ being
the actual tip velocity. This area corresponds 
to the well-known self-similarity solution of
a growing circle in two dimensions and is also the exact
cross-sectional area of the three-dimensional Ivantsov paraboloid
of revolution.  The above mapping is 
presumed to become justified far 
enough away from the tip where the component
of the heat flux along $z$ can be neglected.
Using this mapping and the results of a previous analysis of
two-dimensional anisotropic Laplacian growth 
at constant flux \cite{Almetal}, 
he predicted that the width of the
four dendrite fins 
should increase as the 3/5 power of the distance behind the tip
\cite{Bre}, yielding the expression
\begin{equation}
z= -\frac{3}{5}\left(\frac{\sigma^*}{\sigma_2^*}\right)^{1/3} |x|^{5/3},
\label{finshape}
\end{equation}
where $x$ is the interface coordinate normal to the
$z$-axis in a (010) section of the tip and $\sigma_2^*$ ($\sigma^*$) 
is the tip scaling parameter in two (three) dimensions.
Moreover, the ratio $\sigma_2^*/\sigma^*$ 
becomes independent of anisotropy
for weak anisotropy with $(\sigma_2^*/\sigma^*)^{1/3}\approx 1$.

Our goal in this paper is to use the phase-field method
to obtain a characterization of the dendrite tip morphology 
that is sufficiently detailed and accurate to test 
the above theoretical predictions and to make a critical
comparison with existing benchmark experiments \cite{LaCetal,BisBil}.
The present simulations are based on a novel adaptive-step
diffusion Monte Carlo method \cite{PlaKar1,PlaKar2} that provides an
efficient treatment of the large scale diffusion field away
from the growing structure. Thus this method allows us
to investigate a relatively low undercooling range $\Delta\sim 0.1$
($P_{Iv}\sim 0.03$), in contrast to 
previous simulations that were limited to
Peclet numbers of order unity \cite{KarRap}. 
As it turns out, and this is one of the
main findings in this paper, 
the tip morphology in this Peclet number range is already 
ostensibly independent of anisotropy strength and undercooling,
except for a localized shape distortion at the tip that is 
not experimentally relevant. We are therefore 
able to compare meaningfully this morphology to
existing detailed dendrite shape measurements 
in succinonitrile \cite{LaCetal} and xenon \cite{BisBil}, even though
these measurements were carried out at even smaller $\Delta$.

In the next section, we review the basic equations
of the symmetric model of dendritic growth in three dimensions and briefly 
summarize our numerical methods. The numerical
results are then presented in
section III and discussed in section IV. 
Finally, a summary and conclusions are 
presented in section V.

\section{Basic equations}

We study the standard symmetric model of 
solidification in a pure undercooled melt that assumes
equal thermal diffusivities in the solid and
liquid phases. The basic equations of this model
are given by
\begin{eqnarray}
\partial_tu&=&D\nabla^2 u,\label{e1}\\
v_n&=&D\left(\left.\partial_n u\right|_s
-\left.\partial_n u\right|_l\right),\label{e2}\\
u&=&-d_0\sum_{i=1}^2\left[a({\hat n})
+\partial_{\theta_i}^2a({\hat n})\right]\kappa_i,\label{e3}
\end{eqnarray}
where following common notation, $u=(T-T_m)/(L/c_p)$ is
the scaled temperature field that is zero in equilibrium
and equal to $-\Delta$ in the
liquid far from the interface, $\partial_n u|_l$ ($\partial_n u|_s$)
is the normal gradient of $u$ on the liquid (solid) side of the
interface, $v_n$ is the normal velocity of the interface,
$\theta_i$ are the local angles between
the normal $\hat n$ to the interface and the two local principal
directions on the interface, $\kappa_i$ are the
principal curvatures, $d_0=\gamma_0T_Mc_p/L^2$ is the
microscopic capillary length, and $\gamma(\hat n)=\gamma_0 a(\hat n)$
is the anisotropic surface energy where
\begin{equation}
a(\hat n)\equiv
(1-3\epsilon_4)\left[1-
\frac{4\epsilon_4}{1-3\epsilon_4}(n_x^4+n_y^4+n_z^4)\right]
\label{nonaxi}
\end{equation}
for a material with an underlying cubic symmetry, with the
$x$, $y$, and $z$ axes chosen parallel to the $[100]$ directions.
In writing down the interface condition (\ref{e3}),
we have purposely neglected the effect of interface kinetics
that is believed to be negligible for the low undercooling range
where benchmark shape measurements have been carried out
\cite{LaCetal,BisBil}. Moreover, theoretical shape predictions 
to date \cite{BenBre,Bren2,Bre} have neglected this effect.

We use two different numerical methods.
The first, and main one used here, is the 
phase-field approach \cite{Lan:Ma,ColLev} that allows us
to study the dendrite shape for
the full cubic form of the
surface energy defined by Eq. \ref{nonaxi}. In particular,
we simulate the time-dependent evolution of a
single dendrite arm until steady-state growth is reached.
Since we are primarily interested in a low undercooling regime, 
we use the results of the asymptotic analysis of Karma and Rappel
\cite{KarRap} in order to choose the interface width in the
phase-field model about an order of magnitude smaller than 
$\rho$, and thus much larger than $d_0$ for small $\Delta$. 
This analysis also allows us to choose the model 
parameters so as to make kinetic effects
vanishingly small.
Moreover, we use a recently developed
Monte Carlo algorithm to integrate efficiently 
the diffusion equation in the liquid away from the
growing structure. In this approach,
space is divided into two regions. The first region
consists of the solid plus a thin liquid
layer surrounding the interface where the deterministic phase-field
equations are solved on a cubic lattice with
the same choice of computational parameters 
as in Ref. \cite{KarRap}.  The second 
region is the rest of the liquid 
where the diffusion equation is solved using
an ensemble of random walkers that take progressively larger steps
with increasing distance away from the interface. The algorithm
used to interface the deterministic and 
stochastic solutions of the diffusion equation
in these two regions and to update the walkers 
has been summarized in Ref. \cite{PlaKar1} where it
was used to study the early stage of dendritic
evolution. It has also been exposed
in more details in Ref. \cite{PlaKar2} together with the
results of numerical tests and need not be re-described here.

%-------------------------------------------
\begin{figure}
\centerline{
\psfig{file=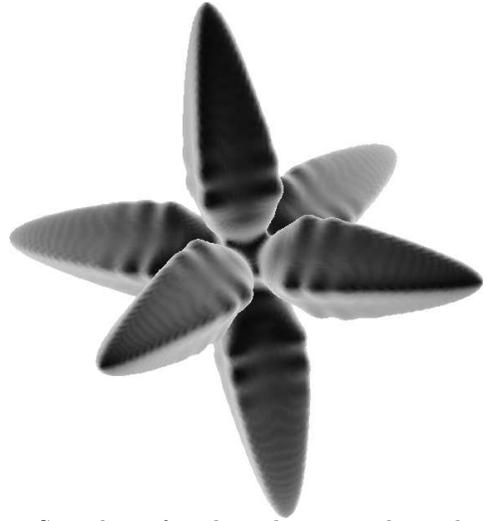,width=.35\textwidth}}
\caption{Snapshot of a three-dimensional simulated 
dendrite with $\Delta=0.1$ and $\epsilon_4=0.025$.}
\label{figsnap}
\end{figure}
%-------------------------------------------
We exploit the cubic symmetry to reduce simulation
time by only integrating a part of a single 
dendrite arm (the domain $x>0$, $0<y<x$ and $z>x$), 
taking advantage of the symmetry planes defined by
$x=0$, $y=0$, $x=y$, and $x=z$. The whole dendrite can
then be reconstructed by successive reflections
at these planes, and the result of one of our
simulations is shown in Fig. \ref{figsnap}.
These simulations were performed on regular
lattices of size $240\times 240 \times 800$
and each took about 200 hours of CPU time on
a 525 MHz DEC-8400 computer. Note that almost
no sidebranches can be discerned, although the
length of the dendrite arms is more than 40 times
the tip radius of curvature. For the analysis
of the steady-state shapes, we used only the
part of the dendrite grown at a constant tip
velocity, which corresponds to about one third
of the arm length in Fig. \ref{figsnap}. To
check whether the shape is well converged, we
performed one run in which the dendrite arm
growing along the positive $z$-direction
was ``cut off'' sufficiently far behind
the tip, and its further evolution was simulated
in a smaller box moving with the dendrite tip
for several diffusion times ($D/V^2$). No
significant change of the shape was observed.
In addition, this run 
(for $\Delta=0.1$ and $\epsilon_4=0.025$) yielded
data which extend considerably farther behind the
tip than for the other parameters, well to within
the range where sidebranches are observed in experiments.

The second numerical method 
is the standard boundary integral
method \cite{KesLev} that can be used to solve 
directly the sharp-interface steady-state
growth equations with the simplified 
axisymmetric form of surface energy
\begin{equation}
a(\hat n)\equiv
(1-3\epsilon_4)\left[1-\frac{4\epsilon_4}{1-3\epsilon_4}
(\cos^4\theta+\frac{3}{4}\sin^4\theta)\right],
\label{axiform}
\end{equation}
which is obtained by averaging the full cubic form (\ref{nonaxi})
over the polar angle $\phi$ in the $x-y$ plane; $\theta$ is
the angle between the local normal to the solid-liquid interface
and the $z$-axis. This method only 
describes axisymmetric tip shapes and is thus only used here
as an additional basis of comparison with phase-field results
regarding the anisotropy-dependent shape distortion in the
tip neighborhood and the selected tip operating state.

\section{Numerical results}

\subsection{Equal cross-sectional area shape}

\begin{figure}
\centerline{
\psfig{file=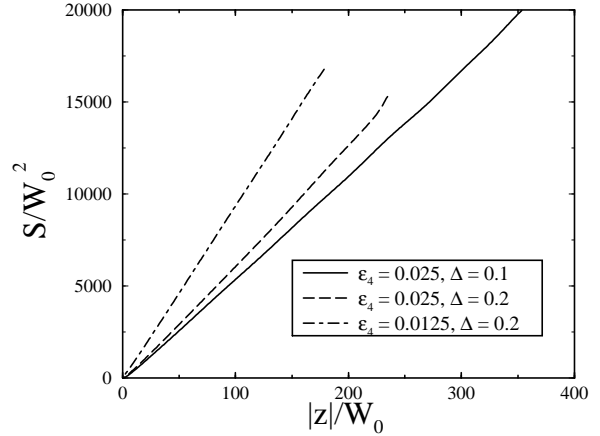,width=.45\textwidth}}
\caption{Plot of the cross-sectional area $S(z)$ vs distance
$|z|$ from the tip for different growth parameters. Lengths
are measured here in units of $W_0$, the thickness of 
the diffuse interfaces in the phase-field model as defined
in Ref. \protect\cite{KarRap}.}
\label{areaplot}
\end{figure}

The cross-sectional area 
of solid normal to the growth axis
was calculated using the formula
\begin{equation}
S(z) = \int {1\over 2}\left(\psi(x,y,z)+1\right)\,dx\,dy,
\label{cross}
\end{equation}
where $\psi(x,y,z)$ is the phase-field that varies from $+1$
in the solid to $-1$ in the liquid in the present model \cite{KarRap}
and $\psi(x,y,z)=0$ defines the solid-liquid interface.
This formula is accurate far enough away from the tip, but of course
not at the tip itself, which suffices for the present purposes.
Plots of $S(z)$ vs $|z|$ are shown in Fig. \ref{areaplot} for
different undercoolings and anisotropy strengths. 
$S(z)$ is accurately fitted by a
straight line sufficiently away from the tip. 
This allows us to define the `parabolic' tip radius, $\rho_p$,
of a paraboloid of revolution with the same
cross-sectional area as the non-axisymmetric phase-field shape.
Such a paraboloid has a cross-sectional area
\begin{equation}
S_p(z) = 2\pi\rho_p (z_0-z) \quad {\rm for} \quad z<z_0,
\label{crosspar}
\end{equation}
where $z_0$ is its tip position.
By fitting the linear part of $S(z)$ away from the tip computed from
Eq. (\ref{cross}) with Eq. (\ref{crosspar}),
we obtain an accurate estimate 
of $\rho_p$ (note that $z_0$ need not coincide with
the tip position of the full non-axisymmetric shape,
see below). The quality of this fit will be illustrated
below in Fig. \ref{fitpol}, where it is compared to a
different method for determining $\rho_p$.

Remarkably, and in good agreement with theoretical expectation
(Eq. \ref{area}), $\rho_p$ coincides to within a numerical
accuracy of 1-2 \% with the Ivantsov tip radius 
$\rho_{Iv}\equiv 2DP_{Iv}/V$ of the isothermal paraboloid growing
with the same tip velocity as the phase-field shape for
the different undercoolings and anisotropies investigated here.
This indicates that $\rho_p\approx \rho_{Iv}$ is potentially
a good scaling parameter for the entire dendrite shape as 
will be confirmed below. Therefore, unless otherwise stated,
all lengths will be rescaled by $\rho_{Iv}$ for the reminder
of this paper. The origin of the $z$-axis is chosen to
coincide with the tip position of the non-axisymmetric
dendrite shape.

%-------------------------------------------
\begin{figure}
\centerline{
\psfig{file=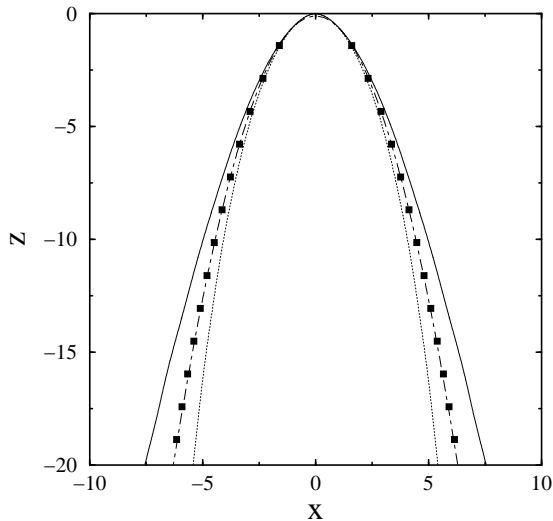,width=.45\textwidth}}
\caption{ 
Sections of phase-field shapes in the
$\phi=0^\circ$ plane (solid line) and $\phi=45^\circ$ plane 
(dotted line), equal cross-sectional 
area shape (solid squares), and parabolic fit
of the latter (dashed line). Parameters are
$\Delta=0.1$ and $\epsilon_4=0.025$.}
\label{longsec}
\end{figure}

%-------------------------------------------
It is also useful to define the equal cross-sectional
area (ECSA) shape
\begin{equation}
r_0(z)\,=\,\left[\frac{1}{2\pi}\int_0^{2\pi}d\phi 
\,r^2(z,\phi)\right]^{1/2}=\left[\frac{S(z)}{\pi}\right]^{1/2}
\label{ecsa}
\end{equation}
where $r(z,\phi)$ is the radial coordinate of the full 
non-axisymmetric shape. This is simply the axisymmetric 
shape that has the same cross-sectional area as the
full shape, and the above results imply at once that
$r_0(z)$ coincides away from the tip with the Ivantsov
paraboloid. The ECSA shape is shown
as solid squares in Fig. \ref{longsec} together with the
parabolic fit obtained by using Eqs. (\ref{crosspar})
and (\ref{ecsa}). The parabolic fit is extended all the way 
to its tip position $z_0$ for illustrative purposes.

%-------------------------------------------
%-------------------------------------------

\subsection{Non-axisymmetric tip morphology}

\begin{figure}
\centerline{
\psfig{file=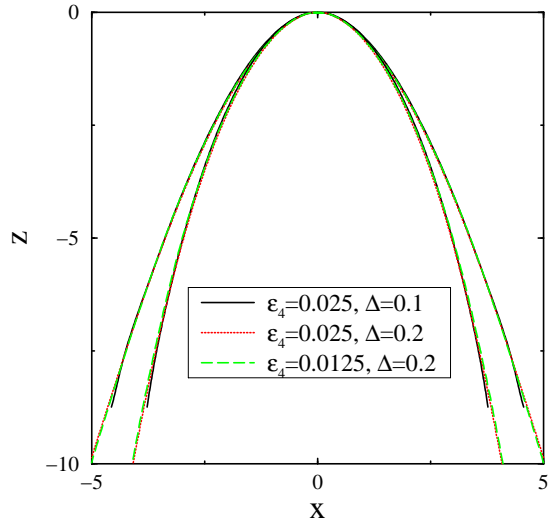,width=.45\textwidth}}
\caption{Superposition of dendrite contours taken along 
fins and valleys for different parameters.}
\label{collapselon}
\end{figure}

\begin{figure}
\centerline{
\psfig{file=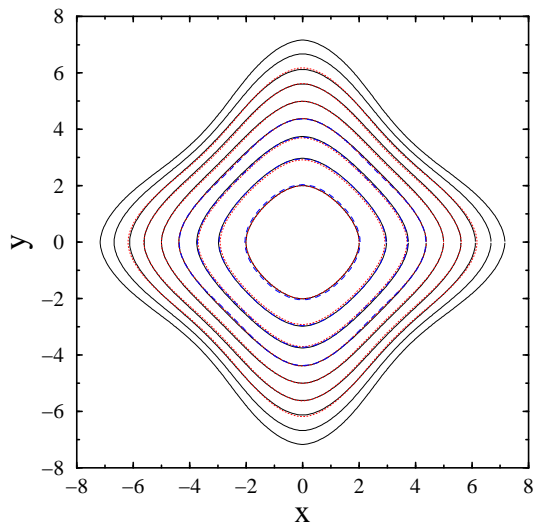,width=.45\textwidth}}
\caption{Superposition of transverse tip sections 
for $\epsilon_4=0.025$, $\Delta=0.1$ (solid lines),
$\epsilon_4=0.025$, $\Delta=0.2$ (dotted lines), and
$\epsilon_4=0.0125$, $\Delta=0.2$ (dashed lines).
Cross-sections are taken at $|z|/\rho_{Iv}=2,4,6,\ldots$.}
\label{collapsetran}
\end{figure}

\begin{figure}
\centerline{
\psfig{file=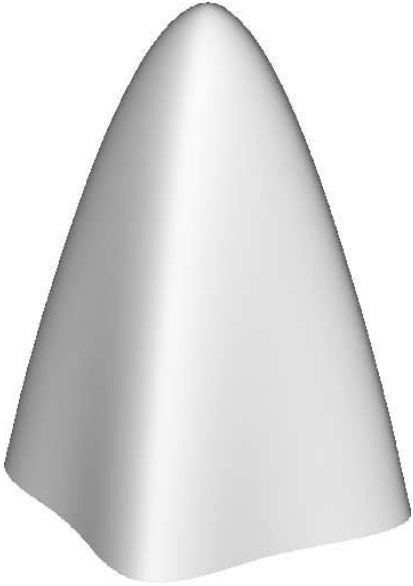,width=.3\textwidth}}
\smallskip
\caption{Three-dimensional view of the simulated tip 
morphology for $\Delta=0.1$ and $\epsilon_4=0.025$.}
\label{3dshape}
\end{figure}

Longitudinal sections of the dendrite tip in the $\phi=0^\circ$
and $\phi=45^\circ$ planes that correspond to the `fins' and 
`valleys' of the non-axisymmetric shape, respectively, 
are superimposed in Fig. \ref{longsec}.
The two contours coincide in the upper region of the tip,
where the shape is essentially axisymmetric, but depart from
a paraboloid (dashed line in Fig. \ref{longsec})
in a small region near the tip that will be examined more closely 
in section III.C. The important point here, especially relevant
for comparisons between simulations and experiments,
is that aside from this localized tip distortion the entire tip
shape scales with $\rho_{Iv}$ and is ostensibly independent of anisotropy
strength and undercooling over the range investigated here. This is
clearly demonstrated by the nearly perfect superposition of longitudinal
and transverse sections of the various tip shapes shown in 
Fig. \ref{collapselon} and Fig. \ref{collapsetran},
respectively. A three-dimensional view of this universal
tip morphology is shown in Fig. \ref{3dshape}.

%-------------------------------------------
%-------------------------------------------

%-------------------------------------------
%-------------------------------------------

%-------------------------------------------
%-------------------------------------------

\begin{figure}
\centerline{
\psfig{file=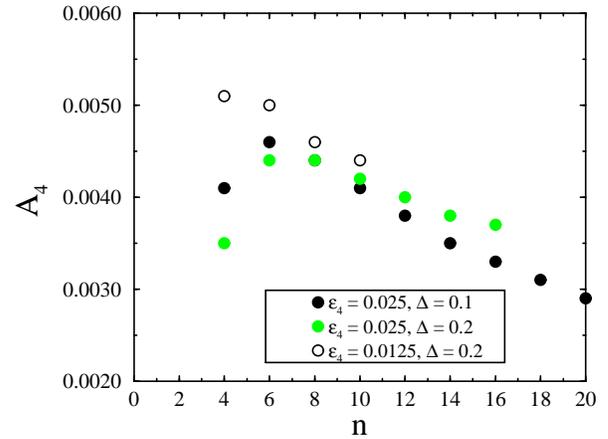,width=.45\textwidth}}
\caption{Plot of $A_4$ vs $n$ obtained from a fourth order polynomial fit
of the fin shape over the interval $\rho_{Iv}\le |z|\le n\rho_{Iv}$.}
\label{fita4}
\end{figure}

\begin{figure}
\centerline{
\psfig{file=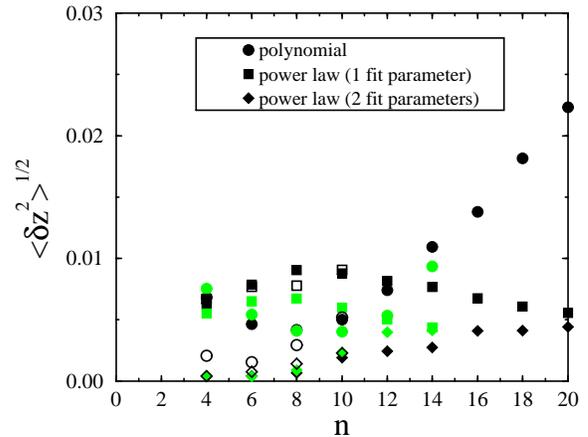,width=.45\textwidth}}
\caption{Plot of the root-mean-square shape deviation corresponding
to the polynomial and power law fits versus the fitting length.
Black symbols: $\epsilon_4=0.025$, $\Delta=0.1$;
grey symbols: $\epsilon_4=0.025$, $\Delta=0.2$;
open symbols: $\epsilon_4=0.0125$, $\Delta=0.2$.}
\label{rms}
\end{figure}

\begin{figure}
\centerline{
\psfig{file=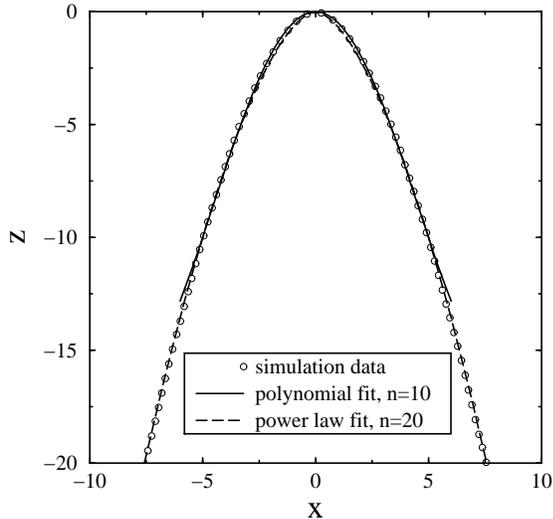,width=.45\textwidth}}
\caption{Comparison of the computed fin shape (open circles)
with the polynomial fit for $n=10$ ($A_4=0.004$, solid line)
and the power law fit for $n=20$ and $\beta=5/3$ 
($a=0.685$, dashed line).}
\label{comparefin}
\end{figure}

In order to test if the analytical
form defined by Eq. \ref{anishape} provides an accurate
description of the non-axisymmetric tip shape, we
determined $A_4$ by minimizing
the spatially averaged root-mean-square (rms)
deviation $<\delta z^2>^{1/2}$,
between the actual shape in the (010)-plane and the 
polynomial form $z=-r^2/2+A_4r^4$ over the 
interval $1\le |z| \le n$
where $n$ was varied from 4 to 20 in steps of 2.
The resulting values of $A_4$ and 
$<\delta z^2>^{1/2}$ are plotted for the different
undercoolings and anisotropy strengths in Fig. \ref{fita4}
and Fig. \ref{rms}, respectively.
Fig. \ref{rms} shows that the form (\ref{anishape}) provides
a good fit of the fin shape for $n$ up to about 10, after
which the rms deviation starts to increase
rather sharply. Fig. \ref{fita4}, in turn, shows that
the fitted value of $A_4$ depends on the fitting
length. This means that the polynomial form (\ref{anishape})
does not exactly characterize the fin shape. However,
the dependence of $A_4$ on $n$ is rather weak:
$A_4$ varies between about $0.004$ 
and $0.005$ for $n$ between 4 and 10. 
A comparison of the polynomial fit and the computed fin 
shape is shown in Fig. \ref{comparefin}, showing an excellent
overlap up to $10\rho_{Iv}$ behind the tip.

\begin{figure}
\centerline{
\psfig{file=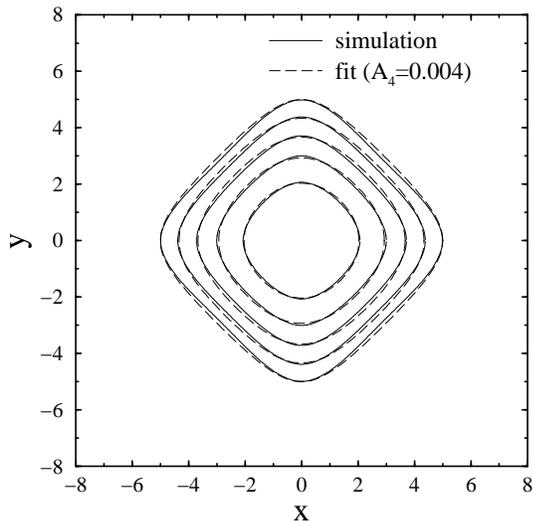,width=.45\textwidth}}
\caption{Comparison of the computed tip cross-sections
for $\epsilon_4=0.025$, $\Delta=0.1$, and the
shape given by Eq. (\protect\ref{anishape}) 
with $A_4=0.004$. Cross-sections are taken at 
$|z|/\rho_{Iv}=2,\,4,\,6,\,8,$ and $10$.}
\label{comparetran}
\end{figure}

We have examined how the form (\ref{anishape}) with
the above procedure to determine $A_4$ fits the
entire non-axisymmetric shape, and not just the fin.
This is illustrated in Fig. \ref{comparetran} where we compare
the transverse sections of the computed phase-field tip shape with
the ones corresponding to the form (\ref{anishape}). 
The two are in good agreement up to ten $\rho_{Iv}$ behind the tip,
although a small deviation due to higher azimuthal harmonics 
is noticeable.

%-------------------------------------------
%-------------------------------------------

%-------------------------------------------
%-------------------------------------------

%-------------------------------------------
%-------------------------------------------

%-------------------------------------------
%-------------------------------------------

\begin{figure}
\centerline{
\psfig{file=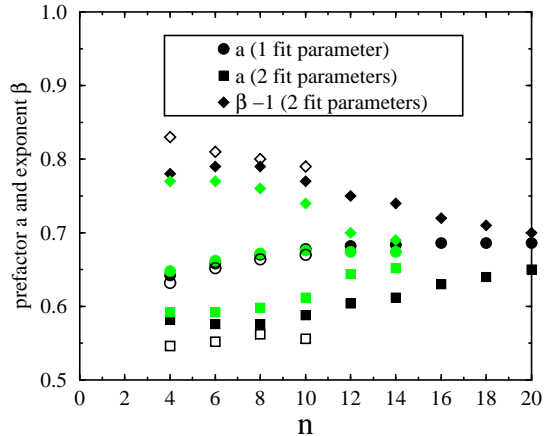,width=.45\textwidth}}
\caption{Plot of the prefactor $a$ and exponent $\beta$ vs
the fitting range parameter $n$.
Black symbols: $\epsilon_4=0.025$, $\Delta=0.1$;
grey symbols: $\epsilon_4=0.025$, $\Delta=0.2$;
open symbols: $\epsilon_4=0.0125$, $\Delta=0.2$.}
\label{abeta}
\end{figure}

We have also fitted the fin shape to the
power law $z=-a|x|^\beta$ by minimizing the 
rms deviation from the computed shape as above. 
This minimization was carried out both with $a$ and
$\beta$ as free parameters and with $a$ as
free parameter and $\beta$ fixed to the theoretically
expected value $5/3$. The values of $a$ and $\beta$
resulting from these two fits are 
plotted vs the fitting range parameter
$n$ in Fig. \ref{abeta} with the corresponding rms deviations
plotted in Fig. \ref{rms}. In the two-parameter fit,
$\beta$ is larger than $5/3$ ($\approx 1.8$) for small $n$ and tends to
$1.7$ for larger $n$, which is close to $5/3$.
Note that the value of $a$ is somewhat smaller
in the two-parameter fit than in the one parameter fit 
because of the larger $\beta$ in the former, and that
the two-parameter fit has a smaller rms deviation as
one would expect. The two fits, however, become 
essentially equally good for large $n$. In this
range, the one-parameter fit yields $a\approx 0.68$ 
independent of anisotropy strength. A comparison 
of the power law fit and the computed fin shape is 
shown in Fig. \ref{comparefin}. In contrast to the
polynomial fit, which becomes inaccurate 
for $|z|>10\,\rho_{Iv}$, the power law closely 
fits the fin shape even far away from the tip.

%-------------------------------------------
%-------------------------------------------

\subsection{Localized tip distortion and tip radius}

\begin{figure}
\centerline{
\psfig{file=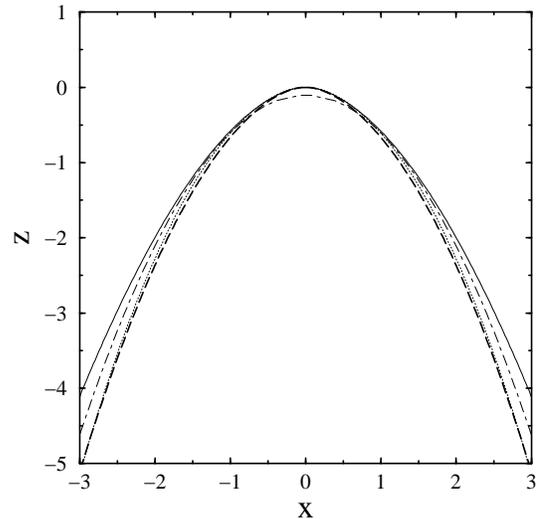,width=.45\textwidth}}
\caption{Magnification of the tip region 
showing the same curves as in Fig. \protect\ref{longsec},
with the solid squares omitted for clarity. Also superimposed
is the axisymmetric shape computed by the boundary integral
method for an azimuthally averaged surface energy 
(thick dashed line). Parameters are $\Delta=0.1$ 
and $\epsilon_4=0.025$.}
\label{tipshape}
\end{figure}

Let us now return to examine more closely the shape
departure from a paraboloid in the region very close to the tip.
For this purpose, we show in Fig. \ref{tipshape} a tip magnification of
the same curves as in Fig. \ref{longsec},
together with the axisymmetric shape 
computed by the boundary integral method
for an azimuthally averaged surface energy.
The latter coincides well with the
fins and valleys in the tip neighborhood, 
but the superposition of these three curves departs
from the parabolic fit of the ECSA shape, which 
represents the interface shape if anisotropic 
capillary effects were absent. Note also that
the boundary integral result starts to deviate 
noticeably from the phase-field shape rather close 
to the tip, approximately at $|z|\approx 0.5\,\rho_{Iv}$.

A quantitative measure of this departure is the
ratio $\rho/\rho_{Iv}$, where $\rho$ is the
actual numerically computed tip radius of the phase-field
or boundary integral shape and $\rho_{Iv}$ 
(recall that $\rho_{Iv}\equiv 2D P_{Iv}/V$ where
$P_{Iv}$ is the Peclet number predicted by Eq. \ref{Ivan}
and $V$ is the numerically computed tip velocity).
For the phase-field shape, $\rho$ is computed from
the $\phi=0^\circ$ section using the interpolation 
scheme that is described in Appendix B of Ref. \cite{KarRap},
and which has been tested against exact boundary
integral benchmark results in two dimensions.

\begin{figure}
\centerline{
\psfig{file=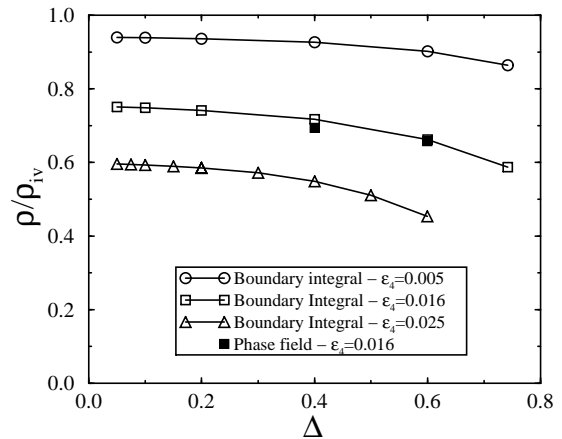,width=.45\textwidth}}
\caption{Ratio $\rho/\rho_{Iv}$ versus
dimensionless undercooling $\Delta$ for different 
anisotropies. Lines are drawn as a guide to the eye.}
\label{rhotipund}
\end{figure}

\begin{figure}
\centerline{
\psfig{file=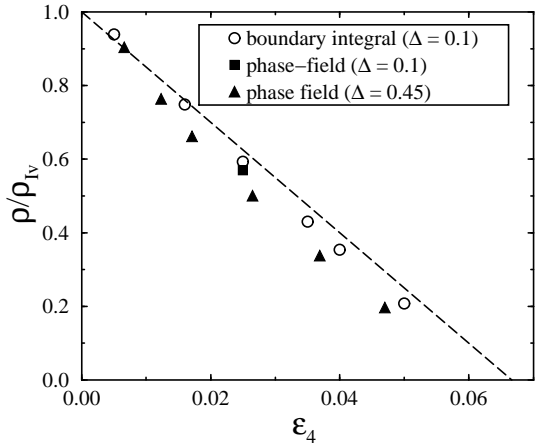,width=.45\textwidth}}
\caption{Ratio $\rho/\rho_{Iv}$ versus
anisotropy for $\Delta=0.1$, and for $\Delta=0.45$
from Ref. \protect\cite{KarRap}.
Superimposed is the constant tip temperature
relation $\rho/\rho_{Iv}=1-15\epsilon_4$ (dashed line).}
\label{rhotipani}
\end{figure}

Figs. \ref{rhotipund} and \ref{rhotipani} show the variation
$\rho/\rho_{Iv}$ as a function of undercooling
and anisotropy, respectively.
The departure of $\rho$ from the Ivantsov relation
increases with anisotropy strength. The same
trend was previously found for a higher undercooling
($\Delta=0.45$) in Ref. \cite{KarRap} 
and the results of these earlier simulations are also shown in
Fig. \ref{rhotipani}. A new finding here is that 
this departure becomes independent of undercooling
in the low undercooling range studied here as can be seen
from the flattening of the curves at small $\Delta$
in Fig. \ref{rhotipund}. For $\Delta\sim 0.1$, $\rho/\rho_{Iv}$ 
is already quasi-independent of $\Delta$ and its small
$\Delta$ limiting value depends solely on the 
anisotropy strength. Interestingly, the variation of
$\rho/\rho_{Iv}$ with anisotropy is crudely approximated 
by the relation
\begin{equation}
\rho/\rho_{Iv}\approx 1-\alpha,\label{tiprad}
\end{equation}
over an order of magnitude change in anisotropy strength,
where $\alpha=15\epsilon_4$ is the stiffness anisotropy.
Since the Gibbs-Thomson condition (\ref{e3}) implies
that the steady-state dendrite tip temperature is given by
\begin{equation}
u_{tip}=-2d_0(1-\alpha)/\rho,
\end{equation}
Eq. \ref{tiprad} is equivalent to stating that the dendrite
tip temperature at a given undercooling is relatively
independent of anisotropy strength.
The straight line corresponding to Eq. \ref{tiprad} is
superimposed as a dashed line in Fig. \ref{rhotipani}.
The numerical results lie slightly
above and below this curve for small and large anisotropy,
respectively.

%-------------------------------------------
%-------------------------------------------

%-------------------------------------------
%-------------------------------------------

%-------------------------------------------
%-------------------------------------------

\section{Discussion}

\subsection{Comparison with analytical theories}

Let us first compare our results with the analytical
theories reviewed in the introduction.
We have found that Eq. \ref{anishape} provides a 
good fit of the tip shape up to a 
distance of about eight to ten $\rho_{Iv}$
behind the tip. Moreover, for low undercooling,
this shape is independent of anisotropy strength (at
least over the range investigated here) which is in
qualitative agreement with the prediction of
linear solvability theory \cite{BenBre,Bren2}.
The value of $A_4$ found here,
however, is about twice smaller than the value
$A_4=1/96\approx 0.104$ \cite{Bren2} predicted by this
theory. One possible reason for this discrepancy
is that the existing solvability calculations 
\cite{BenBre,Bren2} are carried
out in the limit of vanishing anisotropy, whereas
in the present computations the anisotropy is finite.
We have seen, however, that even for a short fitting
distance behind the tip,
$A_4$ increases from about 0.004 to 0.005 when 
the anisotropy is lowered from 2.5\% to 1.25\%, which does
not appear consistent with an extrapolation of
$A_4$ to its theoretically predicted value in the
limit that $\epsilon_4\rightarrow 0$.
It seems also difficult to explain this discrepancy by the fact
that existing calculations are based on linearizing the steady-state
growth equations around a paraboloid of revolution. We find indeed
a localized tip distortion that depends strongly on 
anisotropy strength, and which is not accounted for
in these theories. The rest of the tip shape, however, 
departs only weakly from a paraboloid and is
well fitted by the form (\ref{anishape}) 
independently of anisotropy. Therefore, the origin of
this discrepancy remains to be understood.

In contrast, the predicted power law
(Eq. \ref{finshape}) for the fin shape 
away from the tip is in relatively good quantitative 
agreement with the present simulations. We find a good
fit of the fin shape with a fixed exponent $\beta=5/3$ and a prefactor
$a\approx 0.68$ that is independent of anisotropy strength
and only about 15\% larger than theoretically
predicted, or with a slightly higher exponent $\beta=1.7$ 
and a lower $a\approx 0.65$. It is interesting to note
that the mapping with two-dimensional
growth shapes \cite{Bre} implies that Eq. \ref{finshape} 
should only strictly hold in a region far 
behind the tip where the cross-sectional shape contains
four well-developed arms. It is therefore
rather remarkable, and perhaps coincidental, 
that it also holds closer to the tip in a 
region where the fins are not yet well developed.

Finally, we note that our results indicate a 
smooth cross-over from a polynomial form to a power law
form for the fin shape with increasing $|z|$ as indicated
by the excellent overlap of these two forms in an intermediate 
distance range behind the tip (Fig. \ref{comparefin}).

\subsection{Comparison with experiments}

On the experimental side, the most detailed shape
measurements to date have been carried out by LaCombe {\it et al.} 
who analyzed the full three-dimensional tip morphology
in succinonitrile (SCN) \cite{LaCetal}
and by Bisang and Bilgram who studied
the fin shape in xenon \cite{BisBil}.

\begin{figure}
\centerline{
\psfig{file=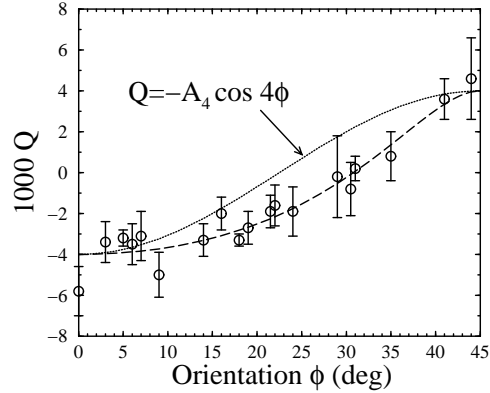,width=.4\textwidth}}
\caption{Function $Q(\phi)$ from the 
experimental measurements of LaCombe {\it et al.} 
in SCN \protect\cite{LaCetal} (circles and error 
bars; some points close to $\phi=0^\circ$ have been 
omitted). The function $Q(\phi)$ computed here from the 
projections of the shape defined by Eq. \protect\ref{anishape} 
with $A_4=0.004$ is shown as a dashed line. This curve
differs from a pure cosine function, which is also shown
for comparison.}
\label{qphi}
\end{figure}

\begin{figure}
\centerline{
\psfig{file=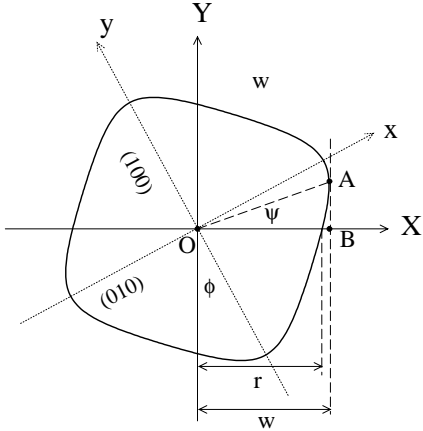,width=.4\textwidth}}
\caption{Sketch illustrating the difference between
the contour of a dendrite as seen under an angle $\phi$
with respect to a (010)-plane and the `true' dendrite
contour at this angle.}
\label{figproj}
\end{figure}

On the basis of their measurements on SCN dendrites,
Lacombe {\it et al.} reported that
the tip shape can be fitted by the form 
\begin{equation}
z\approx -\frac{r^2}{2} - Q(\phi)\, r^4, \label{projshape}
\end{equation}
where lengths are in units of the tip radius. 
The values for the function $Q(\phi)$ reported by
these authors and the associated error bars 
are reproduced here in Fig. \ref{qphi}. It can be seen
that $Q(\phi)$ differs from a cosine function,
and they concluded that the tip morphology cannot be
described by a pure $\cos 4\phi$ mode,
in apparent disagreement with solvability theory and the
present results. This function, however, was constructed
from {\it projected shapes} obtained by taking photographs 
of the dendrite tip from different azimuthal angles $\phi$. 
We point out here that Eq. \ref{projshape} with $Q(\phi)$ 
obtained from projected shapes does not correctly 
represent the actual non-axisymmetric
tip morphology from which these projections are obtained. 
The reason is
that the projected shape observed from a given azimuthal 
angle $\phi$ differs from the cross-sectional shape at 
this angle. To illustrate this point, we have drawn
in Fig. \ref{figproj} a cross-section of the tip shape normal 
to the growth direction and rotated it by an angle $\phi$
with respect to the viewing direction chosen parallel to
the $Y$-axis. The projected shape appears wider than
the `true' contour of the dendrite at that orientation.
Therefore, the dendrite cross-sections (Fig. 11 in Ref. \cite{LaCetal}) 
reconstructed by LaCombe {\it et al.} using
Eq. \ref{projshape} are not representative of the
true cross-sections.

%-------------------------------------------
%-------------------------------------------

%-------------------------------------------
%-------------------------------------------

It is nonetheless possible to relate our results
to the experimental measurements of these authors by calculating 
numerically the function $Q(\phi)$ 
that corresponds to the tip morphology characterized
by a single $\cos 4\phi$ mode (Eq. \ref{anishape}).
As mentioned earlier, Eq. \ref{anishape} 
with $A_4=0.004$ provides a good
fit to our simulated tip morphology up to 10$\rho_{Iv}$ behind the tip 
and is therefore adequate for the purpose of this comparison (neglecting
small corrections due to higher azimuthal harmonics).
To obtain $Q(\phi)$, we must first calculate numerically the
projected shape, and thus relate the {\it apparent}
width $w(\phi,z)$ of the dendrite at a given distance
$z$ behind the tip to the true width $r(-\phi,z)$,
which is the intersection of the fixed 
$X$-axis with the interface.
Using the definitions of 
Fig. \ref{figproj} and purely geometrical considerations,
it is simple to obtain the relation
\begin{equation}
w(\phi,z)=r(\psi-\phi,z)\cos\psi\label{f1}
\end{equation}
where $\psi(\phi,z)$ is defined implicitly by
the relation
\begin{equation}
\psi={\rm arctan}
\left[\frac{\partial_\psi r(\psi-\phi,z)}{r(\psi-\phi,z)}
\right]. \label{f2}
\end{equation}
Eq. \ref{f2} follows directly from the condition
that the point on the true contour that corresponds to
the apparent (perceived) edge of 
the solid tip (point A in Fig. \ref{figproj})
be the one of maximum width
as a function of $\psi$. For a fixed $\phi$, the
corresponding $\psi$ can be
simply obtained by setting equal to zero the 
derivative of the right-hand-side
of Eq. \ref{f1} with respect to $\psi$. 
The angle $\psi$
defined implicitly by Eq. \ref{f2} is then the polar
coordinate of this point measured from the fixed $X$-axis
in Fig. \ref{figproj}, and $\phi-\psi$ measures the angle
between the line OA and the (010) plane. 
Note that at this point, Eqs. \ref{f1} and
\ref{f2} can be used to construct the projected shape
that corresponds to an arbitrary true shape whose
transverse sections are convex.
To proceed further, we now restrict our attention to the case
where the true shape is given by Eq. \ref{anishape}, or
equivalently
\begin{equation}
r(\phi,z)=\left[\frac{1-\sqrt{1-16 z A_4\cos 4\phi}}
{4A_4\cos 4\phi}\right]^{1/2}\label{f3}
\end{equation}
Combining Eqs. \ref{f1} through \ref{f3}, 
we obtain numerically the projected shape 
$z(\phi,w)$ that is the inverse of $w(\phi,z)$ at fixed $\phi$.
We finally obtain $Q(\phi)$ by following
the same procedure as LaCombe {\it et al.}
that consists of fitting these shapes to
a fourth order polynomial of the form
\begin{equation}
z\approx -w^2/2-Q(\phi)w^4 \label{projshapew}
\end{equation}
for different values of $\phi$.
This procedure yields a function $Q(\phi)$ that differs from $\cos 4\phi$
and is superimposed as a dashed line in Fig. \ref{qphi}.
This function fits well through the data points for 
$Q(\phi)$ reported by LaCombe {\it et al.} 
We therefore conclude that, within experimental error
bars, our simulated universal tip morphology 
dominated by a single $\cos\, 4\phi$ mode with an amplitude
$A_4\approx 0.004$ agrees quantitatively well 
with the true underlying tip morphology of SCN dendrites.

A few additional remarks should be made. Firstly,
in drawing the above conclusion we have
implicitly assumed that the inverse problem that consists
of reconstructing the true shape from its projections 
has a unique solution. It can indeed be shown 
that this solution is unique \cite{McFadden}, such that this assumption is
valid. That is if we were to calculate the true
shape from the projections defined by Eq. \ref{projshapew}, we would
recover a tip shape that is well fitted by Eq. \ref{anishape}.
Moreover, an explicit solution to this inverse problem
can be obtained analytically by 
cleverly noting \cite{McFadden} that it
is exactly analogous to the one of constructing
a two-dimensional equilibrium crystal shape,
with the role of the anisotropic surface energy
being played here by $w(\phi,z)$ at fixed $z$. 
This is the well-known Wulff construction \cite{Wood} and
the application of the $\xi$-vector formalism of Cahn and Hoffman
\cite{CahHof} yields that the distance $AB$ in Fig. \ref{figproj} is equal
to $|\partial_\phi w(\phi,z)|$, and thus that
the solution to this inverse problem is given by 
\begin{eqnarray}
r(\psi-\phi,z)&=&\sqrt{[w(\phi,z)]^2+[\partial_\phi w(\phi,z)]^2}
\label{ch1},\\
\psi&=&-{\rm arctan}
\left[\frac{\partial_\phi w(\phi,z)}{w(\phi,z)}\right]
\end{eqnarray}
It should therefore be in principle 
possible to reconstruct the higher azimuthal
harmonics from the experimental data. At present, however,
such an analysis is precluded by the magnitude of the error 
bars in the measurements, especially in view of the small 
amplitude of such harmonics predicted by our simulations.
Lastly, we note that in the present
example the exact (numerically computed) projected shapes
$z(\phi,w)$ defined by Eqs. \ref{f1}-\ref{f3} are 
only approximately fitted by fourth order polynomials in $w$,
even though the underlying true shapes
defined by Eq. \ref{anishape} are exact
fourth order polynomials in $r$. It is simple to see why
this is so by using Eqs. \ref{f1}-\ref{f3} 
to derive an analytical expression for
the projected shape in powers of $A_4$ (valid 
close to the tip), and by noting that this expansion generates 
terms $\sim w^6$ at $O(A_4^2$).
Nonetheless, the form (Eq. \ref{projshapew})
provides a reasonably accurate global fit of the projected shape
over the range $0<|z|<10$ and is therefore
quantitatively adequate to interpret
experimental results.

In xenon, Bisang and Bilgram have reported
that they cannot accurately fit the fin shape 
with a low order polynomial.
They find, instead, a good fit with a power law 
$|z|=a x^\beta$ with $a=0.58\pm 0.04$ and $\beta=1.67\pm 0.05$,
that extends rather close to the tip. Our present results 
differ from theirs in that both for the polynomial
and the power law fit, the fitting parameters vary
with the fitting length, including the exponent
$\beta$ which increases when the fit is restricted
to a region close to the tip. On the other hand, 
in agreement with their findings, the polynomial
fit becomes inaccurate for larger fitting lengths, 
whereas the power law fits the fin shape even far
behind the tip. While the exponents obtained from
our simulations and their measurements are very close,
we find here a slightly larger prefactor $a\approx 0.68$,
independently of anisotropy. Note, however, that
this discrepancy might be simply due to the choice
of the rescaling length. Bilgram and Bisang use the
actual tip radius as a scaling length, whereas we
use the Ivantsov tip radius. Taking this into account,
their prefactor $a'$ and our prefactor $a$ should
be related by $a'=a(\rho/\rho_{Iv})^{\beta-1}$. As
$\rho$ is smaller than $\rho_{Iv}$, our prefactor 
should indeed be larger than theirs. It would be 
interesting to re-examine the experimental data 
using $\rho_{Iv}$ as the scaling length.

\subsection{Tip distortion}

The tip distortion analyzed in section III.C can
be interpreted to result from capillary effects at the tip
that persist even in the limit of low undercooling.
It may appear at first counter-intuitive that such
effects remain important in this limit since the
magnitude of capillary corrections to the tip 
temperature vanishes due to the increase
of the tip radius, i.e. $d_0/\rho\rightarrow 0$ as
$\Delta\rightarrow 0$. One must recall, however, that
the tip velocity and thus the temperature gradient at the tip
also vanishes in this limit. Thus the correct measure
of the relative importance of capillary effects is the
ratio of the magnitudes of the normal gradient of $u$ in the
tip region induced by capillary
variations, $(\partial_n u)_c \sim d_0/\rho^2$, and by heat diffusion,
$(\partial_n u)_d \sim V/D$, or $(\partial_n u)_c /(\partial_n u)_d
\sim Dd_0/(\rho^2V) \sim \sigma^*$. The constancy of $\sigma^*$
implies that capillary effects can produce a non-negligible tip
distortion that persists in the low undercooling limit.

With regard to the tip radius, 
Eq. \ref{tiprad} can be crudely interpreted
to be the simplest linear interpolation between the
isotropic limit where the tip radius approaches its
isothermal value, and the limit $\alpha\rightarrow 1$
($\epsilon_4\rightarrow 1/15$) which marks the
appearance of cusps at the $[100]$ orientations 
of the equilibrium shape where $\rho$ vanishes.
There is, of course, no obvious 
reason why this linear interpolation
should exactly hold in between these two limits and
Fig. \ref{rhotipani} shows that our numerical
results do not lie exactly on it. Eq. \ref{tiprad}
should only be considered a reasonable first estimate of 
the actual tip radius.

\subsection{Operating state}

Traditionally, the operating state of the dendrite 
tip has been characterized in terms of two
independently measurable parameters, the tip radius $\rho$
and velocity $V$, from which one defines 
the scaling parameters
\begin{eqnarray}
P&=&\frac{\rho V}{2D} \label{deff1}\\
\sigma^*&=&\frac{2Dd_0}{\rho^2V} \label{deff2}
\end{eqnarray}
It is clear, however, from the present results
that an accurate measurement of the `true' tip 
radius is most likely not experimentally feasible. Such
a measurement would require a
very high resolution of the localized shape distortion near
the tip that is already barely noticeable on the scale of
Fig. \ref{longsec} for a 2.5\% anisotropy, and with
the interface and the reference parabola represented by
thin lines that are finer than the experimental resolution.

The present results show that, outside the small
region very close to the tip where this distortion is 
noticeable, the entire tip shape is well fitted
by a paraboloid with a small non-axisymmetric four fold 
deviation, and that the rest of the dendrite shape 
further away from the tip scales with
the tip radius $\rho_p$ of this paraboloid.
Therefore, a better definition of the
tip operating state is to use 
$\rho_p$ instead of $\rho$ and to 
define accordingly the dimensionless 
parameters
\begin{eqnarray}
P_p&=&\frac{\rho_pV}{2D}\label{def1}\\
\sigma^*_p&=&\frac{2Dd_0}{\rho_p^2V}\label{def2}
\end{eqnarray}
which are actually the ones that have been traditionally
measured in experiments. In addition, the definition of $\sigma^*$ 
used in solvability theories (see section I)
coincides with the latter definition with the further assumption
that $\rho_p=\rho_{Iv}$.
For a weakly anisotropic material such as SCN 
$\rho$ is not too different from $\rho_p$ such that
the two definition sets (Eqs. \ref{deff1}-\ref{deff2} and
\ref{def1}-\ref{def2}) are roughly equivalent.
In contrast, for a more strongly anisotropic material such as pivalic
acid (PVA), $\rho_p$ can be about two to five times larger than $\rho$,
and concomitantly $\sigma^*_p$ four to twenty-five times
smaller than $\sigma^*$ if we assume that 
$\epsilon_4$ is somewhere in the range 
$0.025-0.05$, where the lower limit has been
measured by Muschol {\it et al.} \cite{Musetal} and
the upper one by Glicksman and Singh \cite{GliSin}.

If we adopt Eqs. \ref{deff1}-\ref{deff2} as a definition
of the tip operating state,
three questions remain to be addressed.
Firstly, what is the best way to measure $\rho_p$ ? 
Secondly, how accurately does linear solvability predict 
the tip operating state as compared to the present
simulations ? Lastly, does $\rho_p$ necessarily 
equal $\rho_{Iv}$, as found here
and assumed in solvability theory ?

%-------------------------------------------
\begin{figure}
\centerline{
\psfig{file=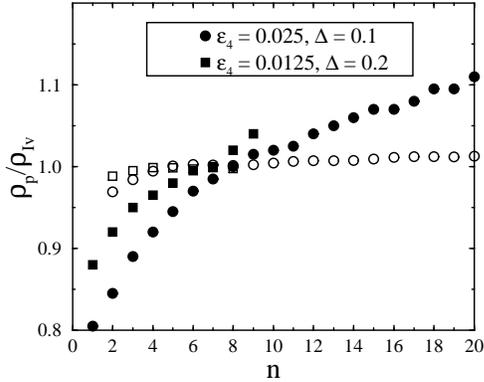,width=.4\textwidth}}
\caption{Plot of the parabolic tip radius $\rho_p$
versus the fitting range $n$ ($|z|<n\rho_{Iv}$) in a fit 
using Eq. \protect\ref{fitpoly} with both $\rho_p$ 
and $A_4$ as fit parameters. For comparison, the
results obtained from a fit of the equal cross-sectional
area shape according to Eq. (\protect\ref{crosspar}) are 
shown as open symbols.}
\label{fitpol}
\end{figure}
%-------------------------------------------

In our simulations, we have obtained $\rho_p$
in section III.A from a plot of the cross-sectional
area of the dendrite versus distance from the tip,
exploiting the fact that, away from the tip, $\rho_p$ is 
simply the slope of this curve divided by $2\pi$.
This provides a very accurate procedure, as can
be demonstrated by plotting the resulting value
for $\rho_p$ versus the fitting range (Fig. \ref{fitpol}):
for $n$ between 4 and 10, $\rho_p$ varies by less than
one percent. The dendrite cross-section, however, is not 
easy to measure as the true dendrite shape is difficult to 
reconstruct from longitudinal projections for
the reasons emphasized in section IV.B. The best data
are obtained for the fin shape \cite{LaCetal,BisBil}. One
way to extract $\rho_p$ is therefore to fit the un-rescaled 
fin shape with a fourth order polynomial,
\begin{equation}
z=-(r/\rho_p)^2/2+A_4(r/\rho_p)^4,\label{fitpoly}
\end{equation}
over a varying distance from the tip
where both $\rho_p$ and $A_4$ are allowed to vary, which
is the method used by LaCombe {\it et al.} in SCN \cite{LaCetal}.
We have carried out this same procedure on 
our computed fin shapes and the results are shown
in Fig. \ref{fitpol} for two different anisotropies. 
(Note that this fit differs from the one carried out in section III.B
where we fixed $\rho_p$ to its value ($\approx\rho_{Iv}$) extracted
from the cross-section measurement, and only varied $A_4$.)
One can see that $\rho_p$ increases with the fitting distance
($=n\rho_{Iv}$) behind the tip, which renders a precise determination
of $\rho_p$ difficult. This trend was observed by Bisang
and Bilgram in xenon dendrites \cite{BisBil}; on the other hand,
for SCN, where $\epsilon_4$ is about twice
smaller than the lowest anisotropy studied here, LaCombe
{\it et al.} find that $\rho_p$ is almost constant
when $n$ varies between 4 and 10. It is therefore possible 
that this fitting procedure improves for lower anisotropies.
Fig. \ref{fitpol} also shows that the departure of the
fitted radius from $\rho_{Iv}$ is of the order of a few
percent for fitting ranges around $n=10$, such
that this method can provide a reasonable estimate.

We compare in Table I the values 
of $\sigma^*_p$ predicted by the linear solvability theory of Barbieri
and Langer \cite{BarLan} with the values corresponding
to the non-axisymmetric shapes of the phase-field model
and the axisymmetric shapes computed by the boundary integral 
method for an azimuthally averaged surface energy.
We list as well the $\sigma^*$ values
corresponding to these shapes. Note that $\rho_p=\rho_{Iv}$ 
both for the paraboloid assumed in solvability theory and for
the computed non-axisymmetric and axisymmetric 
shapes, such that comparing $\sigma^*_p$ values is 
equivalent here to comparing scaled velocity values 
$Vd_0/D=2\sigma^*_pP_{Iv}^2$. One remarkable fact is that
the linear solvability theory predicts relatively accurately
$\sigma^*_p$ even though it does not describe the localized
tip distortion that causes $\rho$ to 
depart from $\rho_p$. Thus, we can conclude from this
comparison that this distortion does not strongly affect 
the selection of the velocity. A reexamination of the 
phase-field results of Ref. \cite{KarRap} in 
terms of $\sigma^*_p$ show, however, that the
linear solvability theory becomes increasingly inaccurate 
for larger anisotropy values ($\epsilon_4>0.03$).

%----------------table-------------------
\begin{table} 
\caption{\label{sstars} Comparison of the selection 
constants $\sigma^*$ obtained
from the present phase-field simulations,
from linear solvability theory, and from
boundary integral calculations.}

\begin{center}
\begin{tabular}{|l|cc|cc|}
\hline
 & $\Delta$ & $\epsilon_4$ & $\sigma^\star$ & $\sigma^\star_p$\\
\hline
Phase-field & $0.2$ & $0.025$ & $0.171$ & $0.0565$\\
Linear Solvability & & & & $0.067$\\
Boundary Integral & & & $0.150$ & $0.055$\\
\hline
Phase-field & $0.1$ & $0.025$ & $0.181$ & $0.0611$\\
Linear Solvability & & & & $0.069$\\
Boundary Integral & & & $0.155$ & $0.059$\\
\hline
Phase-field & $0.2$ & $0.0125$ & $0.0447$ & $0.0282$\\
Linear Solvability & & & & $0.0294$\\
Boundary Integral & & & $0.0394$ & $0.0265$\\
\hline
\end{tabular}
\end{center}
\end{table}
%----------------------------------------

Finally, McFadden {\it et al.} \cite{McFetal} have
recently carried out a perturbative analysis of the diffusion
field around a non-axisymmetric isothermal shape defined 
at leading order in $A_4$ by Eq. \ref{anishape}, 
where $A_4$ is treated as an expansion parameter.
Since their analysis neglects capillary effects,
their tip radius $\rho$ should be compared to $\rho_p$
here. They derived a correction to the
Ivantsov relation (Eq. \ref{Ivan}) that can become
significant in the limit of very low undercoolings.
For the lowest undercooling studied here ($\Delta=0.1$), 
this correction is of the order of a percent and thus
comparable to the accuracy at which $\rho_p$ was
determined numerically. Our results are therefore not in
contradiction with their predictions. We note, however, that
the universal non-isothermal shape found here 
starts to deviate from the isothermal one
they consider only a short distance away from the tip where
the fins develop under the action of anisotropic surface tension.
Simulations at substantially lower undercooling
would be necessary to test if this 
difference between the isothermal and non-isothermal shapes
affects their predictions.

\section{Conclusions}

We have studied the three-dimensional 
morphology of the dendrite tip using recent improvements
of the phase-field method \cite{PlaKar1,PlaKar2,KarRap} 
that make it possible to carry
out quantitatively accurate simulations are relatively
low undercoolings. Our main finding is that
the experimentally measurable
low undercooling tip morphology is independent 
of anisotropy strength and thus universal
under the assumption that kinetic effects are negligible,
in qualitative agreement with solvability theory \cite{BenBre,Bren2}.
The non-axisymmetric deviation of this morphology from
a paraboloid is well fitted by a single $\cos 4\phi$ mode
with an amplitude that is about twice smaller than predicted by 
solvability theory and in good agreement with existing shape
measurements in SCN \cite{LaCetal}. Moreover, these measurements were
reanalyzed here and found to be consistent with 
a non-axisymmetric tip shape dominated
by a single $\cos 4\phi$ mode as in our simulations. 

The fins are well described away from the tip by the
power law derived by Brener on the basis of the analogy \cite{Bre} 
between three-dimensional steady-state shapes and
two-dimensional time-dependent growth shapes \cite{Almetal},
albeit with a slightly larger prefactor than predicted.
Interestingly, the validity of this power law 
extends remarkably close to the tip. Our findings
are also in good agreement with experimental data
on xenon dendrites \cite{BisBil}.

Finally, we conclude that the `true' tip radius is not
an experimentally adequate parameter 
to characterize the tip operating state since the
anisotropy-dependent shape distortion near the tip 
that fixes this radius is most likely not measurable.
In contrast, the tip radius of the paraboloid which
underlies the rest of the tip morphology (excluding this
distortion) is both measurable and a good scaling parameter
for the entire dendrite shape. The latter tip 
radius is indistinguishable from the Ivantsov 
prediction over the range of undercooling
studied here, which does not exclude differences between these
two radii to be present at even lower undercoolings \cite{McFetal}.

It would be interesting in the future to extend the
present study to investigate how the anisotropic kinetics
of molecular attachment at the interface alters the tip morphology. 
The detailed study of the tip morphology, and in
particular its departure
from the universal shape characterized here, may
actually provide a sensitive probe of interface
kinetic effects.

\acknowledgements

This research was supported by US DOE
grant No DE-FG02-92ER45471 and benefited from
computer time allocation at the National Energy
Resource Scientific Computing Center and
the Northeastern University Advanced Scientific
Computation Center. We thank Geoffrey McFadden for
pointing out to us the analogy between the Wulff 
construction and the reconstruction of the dendrite
shape from its projections.


\begin{thebibliography}{999}

\bibitem{Pap} A. Papapetrou, Zeitschrift f\"ur Kristallographie
{\bf 92}, 89 (1935).

\bibitem{Iva}  G. P. Ivantsov, Dokl. Akad. Nauk SSSR {\bf 58}, 567
(1947).

\bibitem{Lan:Hou}  J. S. Langer, in {\it Chance and Matter}, Lectures
on the Theory of Pattern Formation, Les Houches, Session XLVI, edited by
J.  Souletie, J. Vannimenus, and R. Stora (North Holland, Amsterdam,
1987), p.  629-711.

\bibitem{Kesetal:Rev}
D. Kessler, J. Koplik, and H. Levine, Adv. Phys. {\bf 37}, 255 (1988).

\bibitem{BenBre}
M. Ben Amar and E. Brener, Phys. Rev. Lett. {\bf 71}, 589 (1993).

\bibitem{Bren2} E. Brener and V. I. Melnikov, JETP {\bf 80},
341 (1995).

\bibitem{Bre}
E. Brener, Phys. Rev. Lett. {\bf 71}, 3653 (1993).

\bibitem{Almetal} R. Almgren, W. S. Dai, and V. Hakim, Phys. Rev.
Lett.  {\bf 71}, 3461 (1993).

\bibitem{LaCetal}
J. C. LaCombe, M. B. Koss, V. E. Fradkov, and M. E. Glicksman,
Phys. Rev. E{\bf 52}, 2778 (1995).

\bibitem{BisBil} U. Bisang and J. H. Bilgram, Phys. Rev. Lett.
{\bf 21}, 3898 (1995); Phys. Rev. E {\bf 54},5309 (1996).

\bibitem{PlaKar1} M. Plapp and A. Karma, ``Scaling Behavior of
Early Stage Dendritic Growth at Low Undercooling'', 
cond-mat/9906370 (1999).

\bibitem{PlaKar2} M. Plapp and A. Karma, ``Adaptive-Step
Diffusion Monte Carlo Method for Simulating Dendritic Crystal
Growth'', preprint (1999).

\bibitem{KarRap} A. Karma and W.-J. Rappel, 
Phys. Rev. Lett. {\bf 77}, 4050 (1996); Phys. Rev. E
{\bf 57}, 4323 (1998).

\bibitem{Lan:Ma}  J. S. Langer, in {\it Directions in Condensed
Matter}, edited by G. Grinstein and G. Mazenko,
(World Scientific, Singapore, 1986), p. 164.

\bibitem{ColLev} J. B. Collins and H. Levine, Phys. Rev. B {\bf 31},
6119 (1985).

\bibitem{KesLev}
D. A. Kessler and H. Levine,
Acta Metall. {\bf 36}, 2693 (1988).

\bibitem{McFadden} G. B. McFadden (private communications, 1999).

\bibitem{Wood} D. P. Woodruff, {\it The Solid-Liquid Interface}
(Cambridge University Press, London, 1973).

\bibitem{CahHof} J. W. Cahn and D. W. Hoffman, Acta Metall.
{\bf 22}, 1205 (1974).

\bibitem{Musetal}  M. Muschol, D. Liu, and H. Z. Cummins, Phys. Rev. A
{\bf 46}, 1038 (1992).

\bibitem{GliSin}  M. E. Glicksman and N. B. Singh, J. Cryst. Growth
{\bf 98}, 277 (1989).

\bibitem{McFetal} G. B. McFadden, S. R. Coriell, and R. F. Sekerka,
``Analytical Solution for a Non-axisymmetric Isothermal Dendrite'',
preprint (1999).

\bibitem{BarLan}
A. Barbieri and J. S. Langer, Phys. Rev. A{\bf 39}, 5314 (1989).

\end{thebibliography}
\end{document}